# Electromagnetic force distribution inside matter

Masud Mansuripur*, Armis R. Zakharian[†], and Ewan M. Wright*

*College of Optical Sciences, The University of Arizona, Tucson, Arizona 85721
[†]Corning Incorporated, Science and Technology Division, Corning, New York 14831

**Abstract**: Using the Finite Difference Time Domain method, we solve Maxwell's equations numerically and compute the distribution of electromagnetic fields and forces inside material media. The media are generally specified by their dielectric permittivity $\varepsilon(\omega)$ and magnetic permeability $\mu(\omega)$, representing small, transparent dielectric and magnetic objects such as platelets and micro-beads. Using two formulations of the electromagnetic force density, one due to H. A. Lorentz [Collected Papers **2**, 164 (1892)], the other due to A. Einstein and J. Laub [Ann, Phys. **331**, 541 (1908)], we show that the force-density distribution inside a given object can differ substantially between the two formulations. This is remarkable, considering that the total force experienced by the object is always the same, irrespective of whether the Lorentz or the Einstein-Laub formula is employed. The differences between the two formulations should be accessible to measurement in deformable objects.

**1. Introduction**. The classical theory of electrodynamics is based on Maxwell's microscopic equations and the Lorentz law of force [1,2]. Maxwell's equations relate the electric field $\boldsymbol{E}(\boldsymbol{r},t)$ and the magnetic induction $\boldsymbol{B}(\boldsymbol{r},t)$ to the distribution of electric charge $\rho(\boldsymbol{r},t)$ and electric current $\boldsymbol{J}(\boldsymbol{r},t)$ in space-time. In the presence of $\boldsymbol{E}$ and $\boldsymbol{B}$ fields, the Lorentz law specifies the force-density $\boldsymbol{F}$ experienced by material media, which are the seats of $\rho$ and $\boldsymbol{J}$, as follows:

$$\boldsymbol{F}(\boldsymbol{r},t) = \rho(\boldsymbol{r},t)\boldsymbol{E}(\boldsymbol{r},t) + \boldsymbol{J}(\boldsymbol{r},t) \times \boldsymbol{B}(\boldsymbol{r},t). \tag{1}$$

In addition to electric charge and current, material media are also hosts to electric and magnetic dipoles, the density of which are generally represented by polarization $\boldsymbol{P}(\boldsymbol{r},t)$ and magnetization $\boldsymbol{M}(\boldsymbol{r},t)$. In the presence of $\boldsymbol{P}$ and $\boldsymbol{M}$, Maxwell's equations are usually written in their macroscopic form [1-3], in terms of the electric field $\boldsymbol{E}$, the displacement field $\boldsymbol{D} = \varepsilon_o \boldsymbol{E} + \boldsymbol{P}$, the magnetic field $\boldsymbol{H}$, and the magnetic induction $\boldsymbol{B} = \mu_o \boldsymbol{H} + \boldsymbol{M}$. Here $\varepsilon_o$ and $\mu_o$ are, respectively, the permittivity and permeability of free space. (The SI system of units is used throughout the paper.) One may choose to rearrange Maxwell's macroscopic equations in such a way as to eliminate the $\boldsymbol{D}$ and $\boldsymbol{H}$ fields, in which case the resulting equations resemble the microscopic equations with *effective* charge and current densities [1-5] that are given by

$$\rho(\boldsymbol{r},t) = \rho_{\text{free}}(\boldsymbol{r},t) - \nabla \cdot \boldsymbol{P}(\boldsymbol{r},t), \tag{2a}$$

$$\boldsymbol{J}(\boldsymbol{r},t) = \boldsymbol{J}_{\text{free}}(\boldsymbol{r},t) + \partial \boldsymbol{P}(\boldsymbol{r},t)/\partial t + \mu_o^{-1} \nabla \times \boldsymbol{M}(\boldsymbol{r},t). \tag{2b}$$

Applying the Lorentz law to the above charge- and current-density distributions, one obtains the electromagnetic (EM) force-density according to the Lorentz formulation, as follows:

$$\boldsymbol{F}_L(\boldsymbol{r},t) = (\rho_{\text{free}} - \nabla \cdot \boldsymbol{P})\boldsymbol{E} + (\boldsymbol{J}_{\text{free}} + \partial \boldsymbol{P}/\partial t + \mu_o^{-1} \nabla \times \boldsymbol{M}) \times \boldsymbol{B}. \tag{3}$$

Associated with the above force-density is an EM torque-density relative to the origin of coordinates – an arbitrarily chosen reference point. The Lorentz torque-density is given by

$$\boldsymbol{T}_L(\boldsymbol{r},t) = \boldsymbol{r} \times \boldsymbol{F}_L(\boldsymbol{r},t). \tag{4}$$

Other characteristics of the EM field that are intimately tied to the Lorentz force of Eq. (3) are the energy density $\mathcal{E}(\boldsymbol{r},t)$, Poynting vector $\boldsymbol{S}(\boldsymbol{r},t)$, EM momentum density $\boldsymbol{\rho}^{(EM)}(\boldsymbol{r},t)$, and stress tensor $\vec{\vec{\mathcal{T}}}(\boldsymbol{r},t)$. These entities have the following expressions in the Lorentz formulation:

$$\mathcal{E}_L(\boldsymbol{r},t) = \tfrac{1}{2}\varepsilon_o \boldsymbol{E}\cdot\boldsymbol{E} + \tfrac{1}{2}\mu_o^{-1}\boldsymbol{B}\cdot\boldsymbol{B}, \tag{5}$$

$$\boldsymbol{S}_L(\boldsymbol{r},t) = \mu_o^{-1}\boldsymbol{E}(\boldsymbol{r},t)\times\boldsymbol{B}(\boldsymbol{r},t), \tag{6}$$

$$\boldsymbol{\rho}_L^{(EM)}(\boldsymbol{r},t) = \varepsilon_o \boldsymbol{E}(\boldsymbol{r},t)\times\boldsymbol{B}(\boldsymbol{r},t), \tag{7}$$

$$\vec{\vec{\mathcal{T}}}_L(\boldsymbol{r},t) = \tfrac{1}{2}(\varepsilon_o \boldsymbol{E}\cdot\boldsymbol{E} + \mu_o^{-1}\boldsymbol{B}\cdot\boldsymbol{B})\vec{\vec{I}} - \varepsilon_o \boldsymbol{E}\boldsymbol{E} - \mu_o^{-1}\boldsymbol{B}\boldsymbol{B}. \tag{8}$$

In Eq. (8) above, $\vec{\vec{I}}$ is the identity tensor. The various characteristics of the field are interconnected through the energy and momentum continuity equations, namely,

$$\boldsymbol{\nabla}\cdot\boldsymbol{S} + \boldsymbol{E}\cdot\boldsymbol{J} + \partial\mathcal{E}/\partial t = 0, \tag{9}$$

$$\vec{\vec{\nabla}}\cdot\vec{\vec{\mathcal{T}}} + \boldsymbol{F} + \partial\boldsymbol{\rho}^{(EM)}/\partial t = 0. \tag{10}$$

In the mid-1960s, Shockley discovered a violation of momentum conservation in electromagnetic systems under certain circumstances, and proceeded to coin the term "hidden momentum" to account for the imbalance [6]. Physical arguments were subsequently advanced to explain the nature of hidden momentum [7-12]. The subject remains somewhat controversial to this day, and the existence and properties of hidden entities continue to be debated [13-22]. Suffice it to say that the action of the *E*-field on magnetization is believed to produce a hidden energy flux $\mu_o^{-1}\boldsymbol{M}(\boldsymbol{r},t)\times\boldsymbol{E}(\boldsymbol{r},t)$ and a hidden momentum density $\varepsilon_o\boldsymbol{M}(\boldsymbol{r},t)\times\boldsymbol{E}(\boldsymbol{r},t)$. Whenever the Poynting vector of Eq. (6) indicates an imbalance or discontinuity in the rate of flow of EM energy, one must recognize the hidden energy flux as the source of the discrepancy. Similarly, whenever the Lorentz force of Eq. (3) or the torque of Eq. (4) are found to produce no changes in the linear or angular momentum of a material system, the incongruity can be resolved by accounting for the time-rate-of-change of the hidden momentum.

While the above approach to electrodynamics is logically consistent, one might ask whether an alternative formulation exists that avoids the need for hidden entities inside magnetic materials. As it turns out, Einstein and Laub proposed such a formulation in 1908 [23], but it appears that, by the time of Shockley's discovery, their contribution had been largely forgotten. It also did not help that Einstein himself, in response to a 15 June 1918 letter from Walter Dällenbach concerning the EM stress-energy tensor, wrote: *"It has long been known that the values I had derived with Laub at the time are wrong; Abraham, in particular, was the one who presented this in a thorough paper. The correct strain tensor has incidentally already been pointed out by Minkowski"* [24]. We now know, however, that the major difference between the Lorentz and Einstein-Laub formulations is the lack of hidden entities inside magnetic materials in the latter. In other words, it can be shown that the *total* force and *total* torque exerted by EM fields on any object are precisely the same in the two formulations, provided that the contributions of hidden momentum to the Lorentz force and torque on magnetic matter are subtracted [5,25,26]. Since the vast majority of the experimental tests of the Lorentz force law pertain to total force and/or total torque experienced by rigid bodies, these experiments can be said to equally confirm the validity of the Einstein-Laub formulation.

It may thus appear that the choice between the Lorentz and Einstein-Laub formulations is a matter of taste; those who feel comfortable with hidden entities may continue to use the Lorentz law, while others can resort to the Einstein-Laub formalism in order to avoid keeping track of hidden entities inside magnetic materials. This apparent equivalence, however, does not stand up to further scrutiny. Even after subtracting the hidden momentum contribution from the Lorentz force, the corresponding force-density *distribution* within an object turns out to be substantially different from that predicted by the Einstein-Laub formulation. We believe that such differences are measurable and, in fact, the scant experimental evidence presently available seems to favor the Einstein-Laub formulation. In the following sections, we use computer simulations to illustrate some of the differences in the force-density distribution between the two formulations. Before presenting our numerical results, however, it will be useful to briefly review the Einstein-Laub formalism.

**2. The Electrodynamics of Einstein and Laub**. A particular arrangement of Maxwell's macroscopic equations eliminates $D$ and $B$, leaving the remaining fields ($E$ and $H$) related to effective *electric* charge and current densities, ($\rho_{\text{free}} - \nabla \cdot P$, $J_{\text{free}} + \partial P/\partial t$), in addition to effective *magnetic* charge and current densities, ($-\nabla \cdot M$, $\partial M/\partial t$). The corresponding force-density equation in this case was given by Einstein and Laub [23] as follows:

$$F_{EL}(r,t) = \rho_{\text{free}} E + J_{\text{free}} \times \mu_o H + (P \cdot \nabla) E + (\partial P/\partial t) \times \mu_o H + (M \cdot \nabla) H - (\partial M/\partial t) \times \varepsilon_o E. \quad (11)$$

The torque-density in the Einstein-Laub formalism can be shown to require three terms [25,26], namely,

$$T_{EL}(r,t) = r \times F_{EL}(r,t) + P(r,t) \times E(r,t) + M(r,t) \times H(r,t). \quad (12)$$

Although Einstein and Laub briefly mentioned the $P \times E$ and $M \times H$ terms of the above expression in the context of birefringent media [23], it is not difficult to prove the need for the inclusion of these terms in Eq.(12) under *all* circumstances. As a simple example, consider a permanently polarized solid cylinder, having uniform polarization $P_o$ along the cylinder's axis, as shown in Fig.1. In the presence of a constant and uniform external field $E(r,t) = E_o$, the term $(P \cdot \nabla) E$ of Eq.(11) does not produce any forces on the cylinder. Therefore, in addition to the $r \times F_{EL}$ term of Eq.(12), one needs the $P \times E$ term to account for the torque experienced by the cylinder. In contrast, the $-(\nabla \cdot P) E$ term of the Lorentz force-density in Eq.(3) produces equal and opposite forces on the cylinder's top and bottom facets in response to the $E$-field $E(r,t) = E_o$. As before, the integrated force on the entire cylinder vanishes, yet Eq.(4) yields the correct torque without needing additional terms. A similar argument justifies the inclusion of the $M \times H$ term in the Einstein-Laub torque-density formula in Eq.(12).

In linear, isotropic media, where $P$ is always parallel to $E$, and $M$ always parallel to $H$, the $P \times E$ and $M \times H$ terms of Eq.(12) automatically vanish. This might explain why Einstein and Laub originally confined the use of these terms to birefringent media. However, aside from justifications based on specific examples, the universality of Eq.(12) can be proven by demonstrating its consistency with the principle of conservation of angular momentum [26].

The energy density, Poynting vector, momentum density, and stress tensor associated with the Einstein-Laub formulation are given by

$$\mathcal{E}_{EL}(\mathbf{r},t) = \tfrac{1}{2}\varepsilon_o \mathbf{E}\cdot\mathbf{E} + \tfrac{1}{2}\mu_o \mathbf{H}\cdot\mathbf{H}, \tag{13}$$

$$\mathbf{S}_{EL}(\mathbf{r},t) = \mathbf{E}(\mathbf{r},t)\times\mathbf{H}(\mathbf{r},t), \tag{14}$$

$$\boldsymbol{\rho}_{EL}^{(EM)}(\mathbf{r},t) = \mathbf{E}(\mathbf{r},t)\times\mathbf{H}(\mathbf{r},t)/c^2, \tag{15}$$

$$\ddot{\vec{\mathcal{J}}}_{EL}(\mathbf{r},t) = \tfrac{1}{2}(\varepsilon_o \mathbf{E}\cdot\mathbf{E} + \mu_o \mathbf{H}\cdot\mathbf{H})\vec{\mathbf{I}} - \mathbf{D}\mathbf{E} - \mathbf{B}\mathbf{H}. \tag{16}$$

The Einstein-Laub formulation gives precisely the same *total* force on any given object as does the Lorentz formulation. This is not difficult to see, considering that the stress tensor of Eq.(8) is identical to that in Eq.(16), when both tensors are evaluated on a closed surface located entirely in free space surrounding the object under consideration. Such a surface is generally used to calculate the EM force exerted on an isolated object. It is also possible to show that the *total* EM torque acting on an isolated object is always the same in the two formulations [5,25].

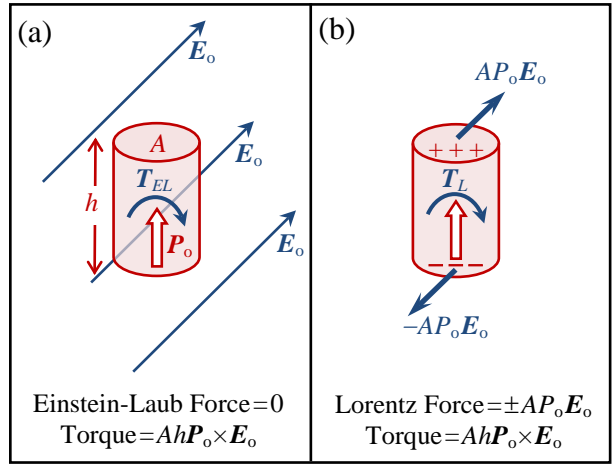

**Fig.1 (Color online)**. A constant and uniform $E$-field $\mathbf{E}_o$ acts on a permanently polarized solid cylinder of cross-sectional area $A$, height $h$, and uniform polarization $\mathbf{P}_o$. (a) In the Einstein-Laub formulation, the force-density $(\mathbf{P}\cdot\nabla)\mathbf{E}$ is zero everywhere and the torque-density is given by $\mathbf{P}_o\times\mathbf{E}_o$. (b) In the Lorentz formulation, bound charges with density $-\nabla\cdot\mathbf{P}$ appear on the top and bottom facets of the cylinder, giving rise to surface charge densities $\pm P_o$. Thus the force exerted by the external $E$-field on these facets is $\pm AP_o\mathbf{E}_o$, even though the net force acting on the cylinder is zero. The torque-density is now given by $\mathbf{r}\times\mathbf{F}_L(\mathbf{r},t)$, which is, once again, equal to $\mathbf{P}_o\times\mathbf{E}_o$.

The astute reader will have noticed that the EM momentum densities in Eqs.(7) and (15) differ by $\varepsilon_o \mathbf{M}\times\mathbf{E}$. This momentum, however, has no observable effects, as it merely balances the hidden momentum of the Lorentz formulation. (The Einstein-Laub formulation, of course, does not have entities hidden inside magnetic matter, neither in the material system nor in the EM fields.) The bottom line is that no measurement of *total* force or *total* torque on a given object can distinguish one formulation from the other.

The distinctive features of the electrodynamics of Lorentz relative to that of Einstein and Laub are in the *distributions* of force and torque throughout material media, which are hosts to $\rho_{\text{free}}$, $\mathbf{J}_{\text{free}}$, $\mathbf{P}$ and $\mathbf{M}$. In other words, despite the equality of total force and total torque, the distributions predicted by the two formulations are not always the same [27,28]. Such distributions should be measurable in experiments on deformable media, such as liquid droplets, and also in conjunction with nonlinear optical phenomena associated with electrostriction or magnetostriction.

The goal of the present paper is to compare and contrast these differing force distributions. We mention in passing that other EM force-density equations due, for example, to Helmholtz, Minkowski, Abraham, and Peierls [8, 29-32], have been proposed and investigated in the past. Reference [29], in particular, contains a wealth of information on Abraham, Minkowski, Einstein-Laub, Helmholtz, and Peierls theories, presenting the relevant theoretical arguments as well as experimental comparisons among the various formulations. Many of the discussions in [29] are, in fact, closely related to the topic of the present paper. The focus of our paper, however, is exclusively on the Lorentz and Einstein-Laub equations, as it is not our goal here to provide an exhaustive comparison of all existing formulations; rather we seek to illustrate, through numerical simulations, the changes in the force-density distribution that could arise when one force equation replaces another – despite the fact that the total force and total torque acting on an isolated object remain the same.

In Minkowski's theory, the force-density only acts where $\nabla \varepsilon$ or $\nabla \mu$ is nonzero (typically at surfaces and interfaces), and there are no forces inside a homogeneous medium. (Here $\varepsilon$ and $\mu$ are the relative permittivity and permeability of the medium.) The Abraham force-density is identical to that of Minkowski in stationary situations. The Abraham and Minkowski force densities, as well as their associated tensors, are in standard use and have been successful in making simple predictions. Again, the total force and torque acting on an isolated body generally agree with those obtained using alternative theories. However, unlike the Einstein-Laub formulation, electrostriction and magnetostriction appear neither in Abraham's nor in Minkowski's theory, and must be introduced separately, resulting in what is commonly known as the Helmholtz force [29]. The Helmholtz force is a description in common use which is similar, but not identical, to the Einstein-Laub force.

Both the Lorentz and Einstein-Laub theories are microscopic (as opposed to phenomenological), and can, in principle, be applied under general circumstances. Both theories allow polarization and magnetization to be related to electric and magnetic fields in nonlinear, nonlocal, anisotropic, dispersive, and hysteretic materials, without hampering one's ability to predict local force and torque densities. In contrast, other formulations may be restricted to linear media, where $\boldsymbol{P}(\boldsymbol{r},t)$ is proportional to $\boldsymbol{E}(\boldsymbol{r},t)$ and $\boldsymbol{M}(\boldsymbol{r},t)$ is proportional to $\boldsymbol{H}(\boldsymbol{r},t)$. [If these linear media also happen to be free from dispersion, the proportionality constants will be the electric and magnetic susceptibilities $\varepsilon_o \chi_e = \varepsilon_o(\varepsilon - 1)$ and $\mu_o \chi_m = \mu_o(\mu - 1)$.]

**3. The Einstein-Laub force inside linear non-absorptive media**. In linear, transparent media, one can express the time-averaged Einstein-Laub force-density in terms of the gradient of the intensity of the EM field. The same cannot be said about the Lorentz force-density, and therein lies a major difference between the two formulations. In the present section we analyze the time-averaged Einstein-Laub force-density in a linear medium specified by its relative permittivity $\varepsilon(\omega)$ and relative permeability $\mu(\omega)$. Here $\omega$ is the angular frequency of a monochromatic, but otherwise arbitrary, EM wave traveling through the medium. The wave induces the following polarization and magnetization at the point $(\boldsymbol{r}, t)$ in space-time:

$$\boldsymbol{P}(\boldsymbol{r},t) = \text{Re}\{\varepsilon_o[\varepsilon(\boldsymbol{r},\omega) - 1]\boldsymbol{E}(\boldsymbol{r},\omega)\exp(-\mathrm{i}\omega t)\}, \tag{17a}$$

$$\boldsymbol{M}(\boldsymbol{r},t) = \text{Re}\{\mu_o[\mu(\boldsymbol{r},\omega) - 1]\boldsymbol{H}(\boldsymbol{r},\omega)\exp(-\mathrm{i}\omega t)\}. \tag{17b}$$

In a non-absorptive medium, where both $\varepsilon(\omega)$ and $\mu(\omega)$ are real-valued, and where $\rho_{\text{free}} = 0$ and $\boldsymbol{J}_{\text{free}} = 0$, the Einstein-Laub force-density of Eq. (11) may be simplified as follows:

$$<F_{EL}(r,t)> = \tfrac{1}{2}\text{Re}\left\{[\varepsilon_o(\varepsilon-1)E\cdot\nabla]E^* - i\omega\varepsilon_o(\varepsilon-1)E\times(B^*-M^*) + [\mu_o(\mu-1)H\cdot\nabla]H^* + i\omega\mu_o(\mu-1)H\times(D^*-P^*)\right\}$$

$$= \tfrac{1}{2}\text{Re}\left\{\varepsilon_o(\varepsilon-1)(E\cdot\nabla)E^* - \varepsilon_o(\varepsilon-1)E\times(-i\omega B)^* + \mu_o(\mu-1)(H\cdot\nabla)H^* + \mu_o(\mu-1)H\times(-i\omega D)^*\right.$$
$$\left. + i\omega\varepsilon_o(\varepsilon-1)E\times\mu_o(\mu-1)H^* - i\omega\mu_o(\mu-1)H\times\varepsilon_o(\varepsilon-1)E^*\right\}$$

$$= \tfrac{1}{2}\text{Re}\left\{\varepsilon_o(\varepsilon-1)(E\cdot\nabla)E^* + \varepsilon_o(\varepsilon-1)E\times(\nabla\times E^*) + \mu_o(\mu-1)(H\cdot\nabla)H^* + \mu_o(\mu-1)H\times(\nabla\times H^*)\right.$$
$$\left. + i\omega\mu_o\varepsilon_o(\varepsilon-1)(\mu-1)(E\times H^* + E^*\times H)\right\}$$

Note that Maxwell's equations $\nabla\times E = -\partial B/\partial t$ and $\nabla\times H = \partial D/\partial t$ have been used in going from the 2nd to the 3rd line of the above derivation. The last term in the final expression is purely imaginary and can therefore be dropped. The remaining terms are then simplified with the aid of the vector identity $\nabla(A\cdot B) = (A\cdot\nabla)B + (B\cdot\nabla)A + A\times(\nabla\times B) + B\times(\nabla\times A)$. We will have

$$<F_{EL}(r,t)> = \tfrac{1}{4}\varepsilon_o[\varepsilon(r,\omega)-1]\nabla(E\cdot E^*) + \tfrac{1}{4}\mu_o[\mu(r,\omega)-1]\nabla(H\cdot H^*). \tag{18}$$

It is thus seen that, in the special case where the EM wave is monochromatic and also $\varepsilon$ and $\mu$ are real-valued, the time-averaged Einstein-Laub force-density consists of two terms, each being proportional to the gradient of the local field intensity. Note that $\varepsilon$ and $\mu$ in Eq. (18) are, in general, functions of the spatial coordinates $r$, yet they remain outside the gradient operator. This proportionality of the Einstein-Laub force-density to the local intensity gradient of the EM field is unique. The examples discussed in the following sections clearly demonstrate that this property is *not* shared by the Lorentz force.

It should be pointed out that certain experimental observations are believed to support the Helmholtz force over that of Einstein and Laub; see, for example, the Hakim-Higham experiment involving the force of static electric fields on liquid dielectrics [33]. Hakim and Higham conclude (as does Brevik [29]) that the Helmholtz theory provides a better fit to their experimental data than does the theory of Einstein and Laub. Interpretation of such experiments, however, as pointed out by Brevik [29], requires careful attention to spurious effects, and, in any case, it is necessary to examine a much broader range of situations before settling on a microscopic theory of EM force and torque that is firmly rooted in physical reality.

**4. Gaussian beam propagating inside a transparent, homogeneous, isotropic dielectric**. With reference to Fig. 2(a), consider a transparent, non-magnetic dielectric of refractive index $n = 2.0$ through which a monochromatic Gaussian beam propagates along the negative $z$-axis. The beam has vacuum wavelength $\lambda_o = 0.65\,\mu\text{m}$, is linearly polarized along $x$, and has a circular cross-section with a full-width-at-half-maximum amplitude FWHM $= 1.5\,\mu\text{m}$. The instantaneous $E$-field profile of the beam is shown in Figs. (2b) and (2c), which are plots of $E_x$ in the $yz$-plane and $E_z$ in the $xz$-plane, respectively. Note that the propagation distance along $z$ is only about $1\,\mu\text{m}$, which is less than the Rayleigh range of the Gaussian beam. This explains why the fraction of the beam shown in these figures remains collimated as it propagates downward along the $z$-axis.

Figure 3 shows plots of the force-density components $F_x$ and $F_y$ in the $xy$ and $yz$ cross-sectional planes of the system depicted in Fig. 2. The Einstein-Laub force-density distribution $F_{EL} = (P\cdot\nabla)E + (\partial P/\partial t)\times\mu_o H$ has been used in these calculations. The left-hand column in Fig. 3 represents the case of an incident beam that is linearly polarized along the $x$-axis, while the right-hand column corresponds to linear polarization along $y$. The arrows overlapping the $F_x$ plots in the $xy$-plane (top row) show the total force-density $F_x\hat{x} + F_y\hat{y}$ in the cross-sectional $xy$-plane.

Clearly, the optical force everywhere tends to compress the host medium toward the $z$-axis. The compressive nature of the force is also apparent in the $F_y$ plots in the $yz$-plane (bottom row).

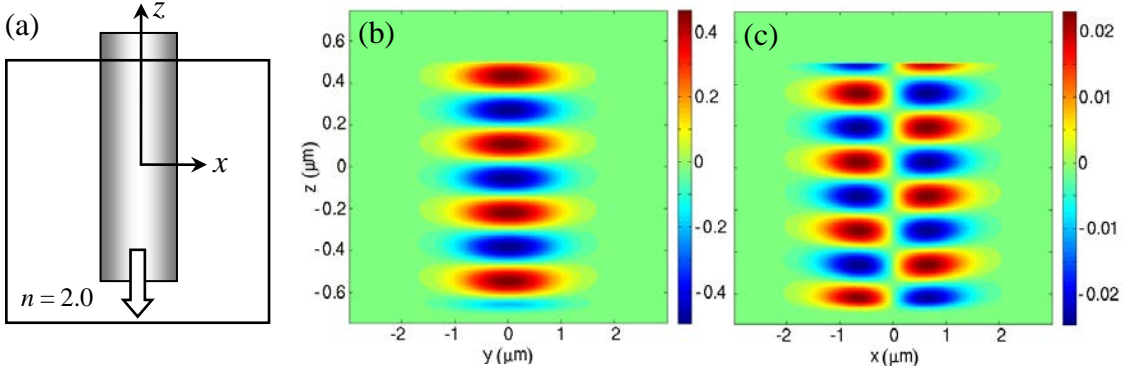

**Fig. 2 (Color online).** (a) A semi-infinite dielectric of refractive index $n = 2.0$ is illuminated from above by a linearly-polarized Gaussian beam having $\lambda_o = 0.65\,\mu\text{m}$ and an amplitude FWHM of $1.5\,\mu\text{m}$. The incident beam, which has a circular cross-section, is linearly-polarized along the $x$-axis. The source is placed inside the dielectric at $z = 0.5\,\mu\text{m}$, and the beam is allowed to propagate downward. Shown in (b) and (c) are plots of instantaneous $E_x$ and $E_z$ in the $yz$ and $xz$ planes, respectively. The integral over the $xy$-plane of the $z$-component of the Poynting vector, $S_z$, yields the total incident optical power as $P_{\text{inc}} = 8.3 \times 10^{-16}\,\text{W}$. This arbitrary value, chosen for numerical convenience, is the scale-factor by which the computed force-density must be normalized.

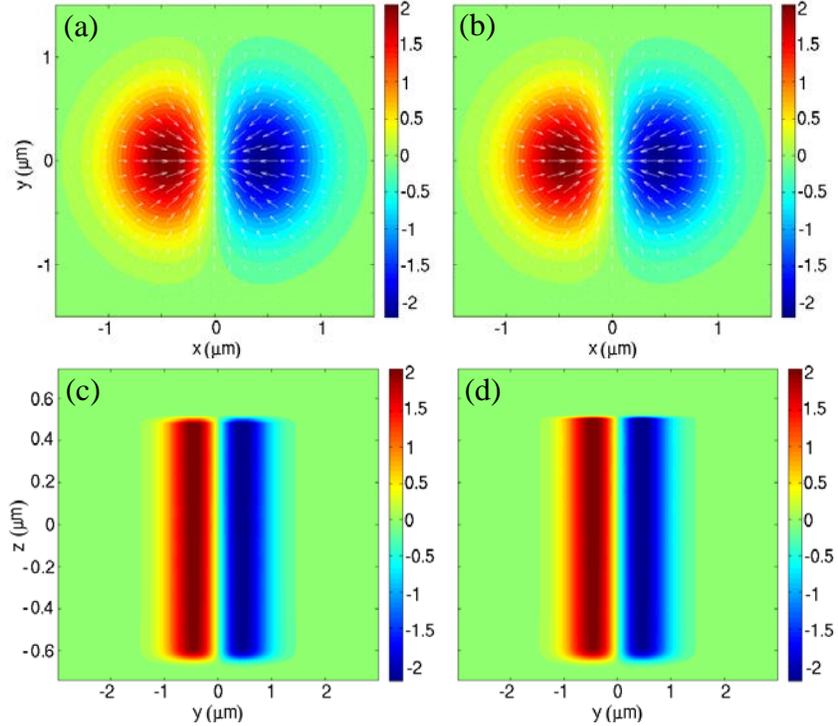

**Fig. 3 (Color online).** Plots of the force-density components $F_x$ and $F_y$ in the $xy$ and $yz$ cross-sectional planes of the system depicted in Fig. 2, computed using the Einstein-Laub formula. The left column represents an incident beam that is linearly polarized along the $x$-axis, while the right column corresponds to linear polarization along $y$. The arrows overlapping the plots of $F_x$ in the $xy$-plane (top row) show the total force density $F_x\hat{\boldsymbol{x}} + F_y\hat{\boldsymbol{y}}$, which tends to compress the host medium radially toward the $z$-axis. The color scale bar shows the force density in units of $\mu\text{N/m}^3$.

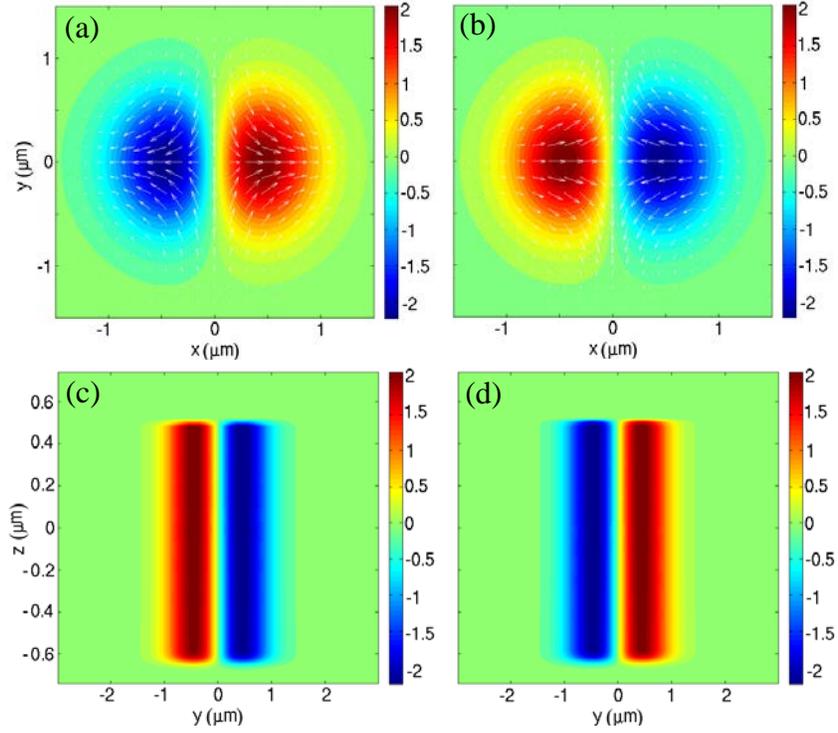

**Fig. 4 (Color online)**. Plots of the Lorentz force-density components $F_x$ and $F_y$ in the $xy$ and $yz$ cross-sectional planes of the system depicted in Fig. 2. The left column represents an incident beam that is linearly polarized along the $x$-axis, while the right column corresponds to linear polarization along $y$. The arrows overlapping the plots of $F_x$ in the $xy$-plane (top row) show the total in-plane force-density $F_x \hat{\boldsymbol{x}} + F_y \hat{\boldsymbol{y}}$, which tends to compress the dielectric host in one direction but expand it in another direction. This behavior of the force is also apparent in the $F_y$ plots in the $yz$-plane (bottom row). The color scale bar shows the computed force density in units of $\mu N/m^3$. This force-density corresponds to an incident optical power $P_{inc} = 8.3 \times 10^{-16}$ W, as pointed out in the caption to Fig. 2.

The Lorentz formula $\boldsymbol{F}_L = -(\boldsymbol{\nabla} \cdot \boldsymbol{P})\boldsymbol{E} + (\partial \boldsymbol{P}/\partial t) \times \boldsymbol{B}$ predicts a very different force-density distribution in the system of Fig. 2, as seen in Fig. 4. According to Figs. 4(a,c), when the $E$-field of the incident beam is parallel to the $x$-axis, the radiation force in the $yz$-plane is compressive (i.e., toward the $z$-axis), while that in the $xz$-plane is expansive. Conversely, Figs. 4(b,d) indicate that, when the incident $E$-field is parallel to $y$, the force in the $xz$-plane is compressive, while that in the $yz$-plane is expansive.

In 1973, Ashkin and Dziedzic performed a remarkable experiment in which they focused a green laser beam ($\lambda_o = 0.53\,\mu m$) onto the surface of pure water [34]. They observed a bulge on the surface, where the focused laser beam had entered. Subsequent analysis by Loudon [35,36] showed that compressive radiation forces beneath the surface tend to squeeze the liquid toward the optical axis, causing a surface bulge via the so-called "toothpaste tube" effect. In his analysis, Loudon used the Einstein-Laub formulation; his findings are consistent with the results of our simulations depicted in Fig. 3, which indicate a compressive force pointing everywhere toward the $z$-axis. In contrast, a theoretical analysis based on the Lorentz formulation [37] revealed the existence of both expansive and compressive forces in different regions beneath the surface, which effectively cancel out, thus ruling out the possibility of bulge formation on the water surface. The simulation results shown in Fig. 4 are in agreement with the latter analysis. The observations of Ashkin and Dziedzic in [34] thus provide a rare experimental evidence against the Lorentz formulation and in support of the Einstein-Laub force-density expression.

One must not forget, however, that the Abraham and Minkowski force densities also predict a bulge similar to that observed in the experiment, although, in this case, electrostriction (i.e., Helmholtz force) is required to account for the "squeeze" of the liquid needed for stability. This alternative explanation of the observed bulge, discussed at length in [29], is qualitatively similar to Loudon's analysis based on the Einstein-Laub equation. Either way, it is clear that the experimental results hint at a departure from the Lorentz law, suggesting the need for further analysis to pinpoint the correct form of the microscopic force equation, one that can accurately predict the measurable characteristics of the bulge in addition to explaining other relevant observations.

**5. Gaussian beam passing through a transparent slab.** The conclusions reached in the preceding section remain valid for a thin dielectric slab of a transparent material as well. Shown in Fig. 5 are the various cross-sectional profiles of the Einstein-Laub force-density distribution in a 2 µm-thick slab of refractive index $n=1.5$. The incident beam is Gaussian and linearly-polarized along the $x$-axis, having $\lambda_o=0.65$ µm and a circular cross-section with FWHM = 1.0 µm. The incident beam, whose optical power (in the air) is $P_{inc}=7.3\times10^{-16}$ W, propagates along the negative $z$-direction. As before, the incident optical power has an arbitrary value chosen for numerical convenience. The computed force densities must subsequently be normalized by the above $P_{inc}$ to yield the force-density per watt of incident optical power.

In Fig. 5, the standing wave created between the entrance and exit facets of the slab is clearly visible in the plotted force profiles in the $xz$ and $yz$ planes. While $F_z$ alternates in direction (pointing up or down depending on the location within a fringe), the lateral component $F_x\hat{x}+F_y\hat{y}$ of the force is radially directed and points inward everywhere, tending to compress the host medium uniformly toward the $z$-axis. In contrast, the Lorentz force-density distribution depicted in Fig. 6 for the same slab under identical conditions shows a compressive lateral force in the $yz$-plane, but an expansive lateral force in the $xz$-plane.

A possible experiment on a thin dielectric slab to distinguish the behavior predicted by the Einstein-Laub equation (Fig. 5) from that suggested by the Lorentz law (Fig. 6) could involve a $Si_3N_4$ membrane having a thickness of only a few microns and supported on a silicon substrate. A small hole etched into the substrate would allow the $Si_3N_4$ membrane to act as a free-standing slab. Silicon nitride would be a good choice for such an experiment as it has minimal inhomogeneity and no crystalline anisotropy that would otherwise complicate the interpretation of the results. More importantly, $Si_3N_4$ is highly transparent at $\lambda \sim 0.8$ µm, where femtosecond pulsed Ti:Sapphire lasers operate. A 150 fs – 1.0 µJ light pulse focused to a diameter of ~10 µm at normal incidence will excite an elastic wave within the membrane. The evolution of this wave, monitored by a local probe (e.g., micro-Raman probe [38]), should readily distinguish between the two force-density formulations. Needless to say, the aforementioned numbers are exemplary; one could, for instance, increase the spot diameter to 50 µm and raise the optical pulse energy to 25 µJ without affecting the following discussion.

With a 6 MW laser pulse focused to an Airy disk diameter of ~10 µm on the surface of the $Si_3N_4$ membrane, the pressure on the cylindrical boundary of the beam (as it passes through the slab) will be $¼\varepsilon_o(n^2−1)|E|^2 \approx 200$ MPa. Considering that Young's modulus of $Si_3N_4$ (depending on the material fabrication method) is somewhere in the range of 150–300 GPa, the 10 µm diameter of the illuminated region should contract by about 0.1% under the action of the 6MW light pulse. The sensitivity of the Raman micro-probe is ~25MPa for silicon [38]. If we assume that $Si_3N_4$ is not too different from silicon in its Raman signal sensitivity, the optically-induced stress will be about an order of magnitude greater than that needed for micro-Raman measurements. (Of course, the stress is induced in the medium only for a short duration after the application of the pump pulse, so averaging over many pulses is needed to obtain a reasonably strong Raman signal.)

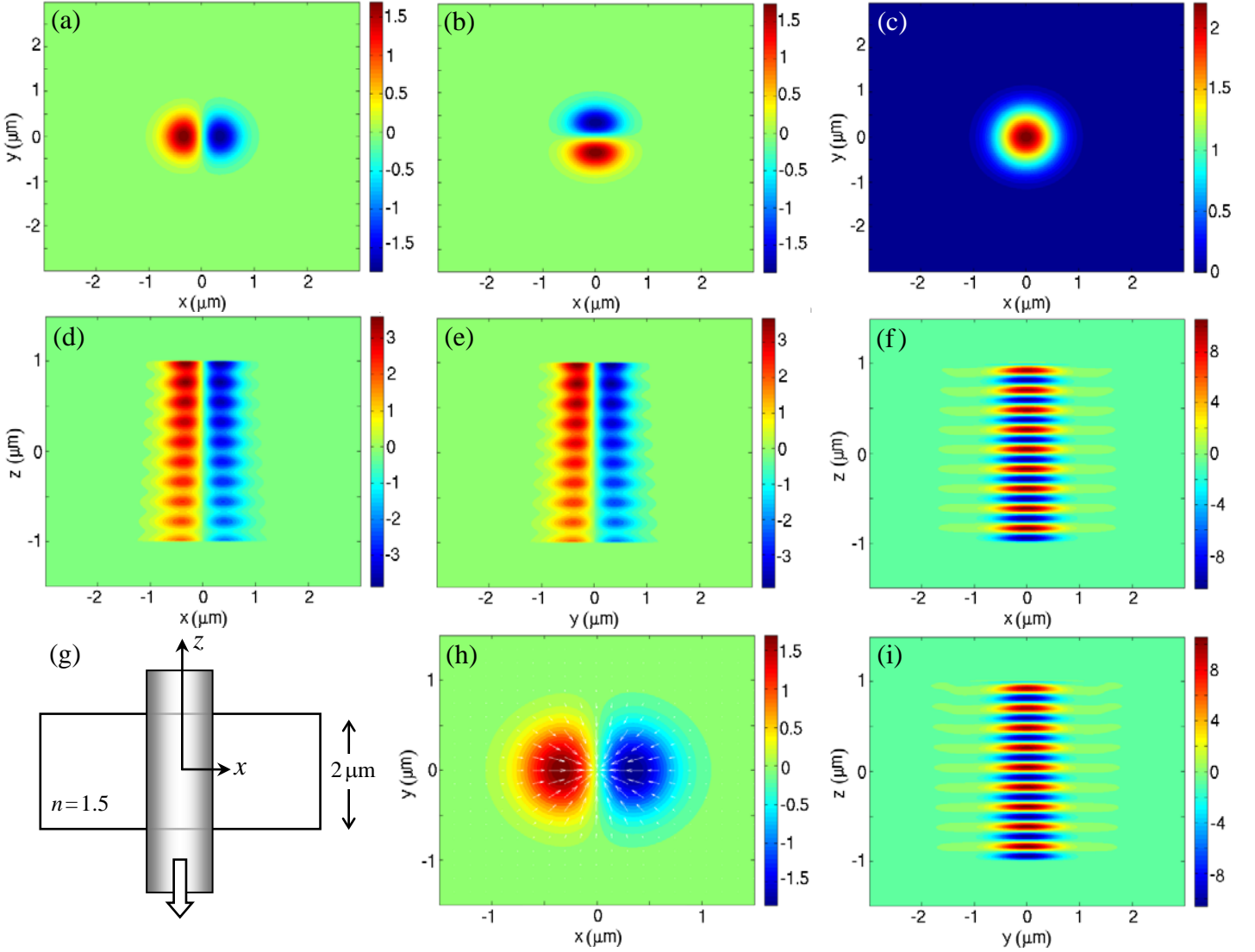

**Fig.5 (Color online).** Plots of the time-averaged Einstein-Laub force-density in various cross-sections through the center of the computational domain. A Gaussian beam of wavelength $\lambda_o = 0.65\,\mu m$, having amplitude FWHM $= 1\,\mu m$, is incident from the air on a $2\,\mu m$-thick dielectric slab of refractive index $n = 1.5$, which occupies the range $z = -1\,\mu m$ to $z = +1\,\mu m$; the source is at $z = 1.2\,\mu m$. The color scale-bars show the computed force-density in $\mu N/m^3$, corresponding to an incident optical power $P_{inc} = 7.3 \times 10^{-16}\,W$. (a-c) $F_x$, $F_y$ and $F_z$ in the central $xy$-plane. (d) $F_x$ in the central $xz$-plane. (e) $F_y$ in the central $yz$-plane. (f) $F_z$ in the central $xz$-plane. (g) Diagram showing the Gaussian beam and the dielectric slab. (h) The vector-field $F_x\hat{x} + F_y\hat{y}$ superposed over the distribution of $F_x$ in the central $xy$-plane. (i) $F_z$ in the central $yz$-plane.

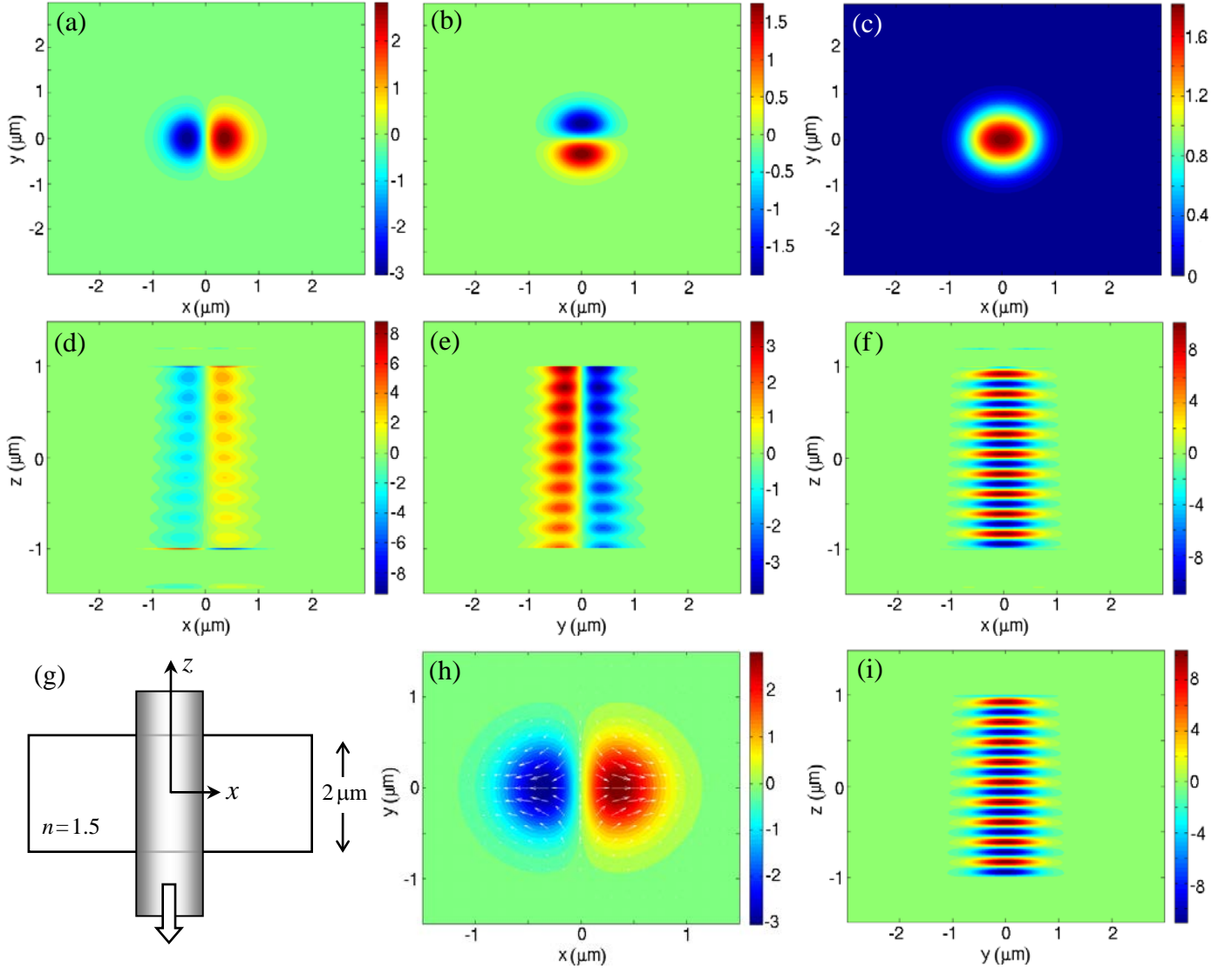

**Fig.6 (Color online).** Plots of the time-averaged Lorentz force-density in various cross-sections through the center of the computational domain. A Gaussian beam of wavelength $\lambda_o = 0.65\,\mu m$, having amplitude FWHM = 1 μm, is incident from the air on a 2 μm-thick dielectric slab of refractive index $n = 1.5$, which occupies the range $z = -1\,\mu m$ to $z = +1\,\mu m$; the source is at $z = +1.2\,\mu m$. The color scale-bars show the computed force-density in $\mu N/m^3$, corresponding to an incident optical power $P_{inc} = 7.3 \times 10^{-16}\,W$. (a-c) $F_x$, $F_y$ and $F_z$ in the central $xy$-plane. (d) $F_x$ in the central $xz$-plane. (e) $F_y$ in the central $yz$-plane. (f) $F_z$ in the central $xz$-plane. (g) Diagram showing the Gaussian beam and the dielectric slab. (h) The vector-field $F_x\hat{x} + F_y\hat{y}$ superposed over the distribution of $F_x$ in the central $xy$-plane. (i) $F_z$ in the central $yz$-plane.

The threshold for ablation due to multi-photon absorption in BK7 glass has been reported at 5.6 μJ for a 150 fs Ti:Sapphire pulse focused to a 10 μm-diameter spot [39]. Assuming that $Si_3N_4$ is not too different in this respect from BK7 glass, our suggested pulse energy of 1.0 μJ over a 10 μm spot is substantially below the ablation threshold. In addition, the extinction coefficient of $Si_3N_4$ at $\lambda = 0.8\,\mu m$ is essentially "zero" according to available literature. It is easy to estimate that the imaginary part of the refractive index must be below $10^{-4}$ for the thermal effects to be negligible. (Assumptions: 1.0 μJ pulse focused to a diameter of 10 μm on a 5 μm-thick film, maximum temperature rise ~10°C).

The relative change in the refractive index with pressure is defined as $(1/n)dn/dP$. Typical numbers for Si, diamond, Boron Nitride, and SiC range from $10^{-3}$ to $10^{-4}$ $GPa^{-1}$. Considering that the radial radiation pressure inside the $Si_3N_4$ membrane is on the order of 0.2 GPa, the fractional change of the refractive index is expected to be in the $10^{-4}-10^{-5}$ range. Placing the sample in a high-Q cavity and monitoring the cavity's transmission would be one way to measure this level of $\Delta n$, although this small value of $\Delta n$ may be easily obscured by a few degrees of temperature rise in the sample. Moreover, the nonlinear refractive index of $Si_3N_4$ should be significant at these intensities. Assuming a typical nonlinear coefficient $n_2 = 10^{-15} cm^2/W$, our intensity of $6\times10^{12}$ $W/cm^2$ will change the refractive index of the illuminated spot by as much as 0.006, swamping the desired signal.

The existence of a nonlinear signal, however, is not necessarily counterproductive, considering that the stress that needs to be monitored is itself a source of nonlinearity (via electrostriction), as discussed in the following section. While the electronic contribution to the Kerr nonlinearity is essentially instantaneous, the electrostrictive nonlinearity has a slower temporal response. The bottom line is that, with a speed of sound around $10^4$ m/s, one picosecond after the high-power pulse has passed through the membrane, the elastic deformations in the area under illumination will have propagated by only ~10 nm. Therefore, with a probe that is delayed by ~1.0 ps relative to the high-power pump pulse, one should be able to monitor the change in the refractive index caused by electrostriction alone.

**6. Contribution of electromagnetic force to non-linear optical effects**. The EM force-density acquires a direct physical meaning in the context of electrostriction of a dielectric medium in the presence of an optical wave. In particular, electrostriction induced by a spatially inhomogeneous E-field causes the host medium to constrict in regions of higher electric field, thereby creating a pressure and mass-density variation throughout the medium. For a broad class of isotropic dielectric media, the acoustic wave equation that governs mass-density variations $\tilde{\rho}(\boldsymbol{r},t)$ in the medium [40,41] is given by

$$\left(\frac{\partial^2}{\partial t^2} - \Gamma'\nabla^2\frac{\partial}{\partial t} - v^2\nabla^2\right)\tilde{\rho}(\boldsymbol{r},t) = -\nabla\cdot\boldsymbol{F}(\boldsymbol{r},t). \qquad (19)$$

Here we have followed the notation of [41], where the mean-density $\rho_0 >> |\tilde{\rho}|$, $v$ is the speed of sound in the host medium, $\Gamma'$ is the damping parameter, and $\boldsymbol{F}$ is the EM force density. (We have changed the sign of the right-hand-side of Eq.(19) compared to Ref. [41] on the basis that the force-density there should have been $\boldsymbol{F} = -\nabla p_{st}$, with $p_{st}$ the strictive pressure, to yield a compressive force-density associated with regions of negative pressure.) The standard theory of electrostriction, derived using thermodynamic arguments and the Lorentz-Lorenz law, yields

$$\boldsymbol{F}_{st} = \tfrac{1}{2}\varepsilon_o[(n^2-1)(n^2+2)/3]\nabla<\tilde{E}^2>. \qquad (20)$$

In the above equation, $\tilde{E}$ is the electric field amplitude of an EM wave propagating in a dielectric medium of refractive index $n$, and the angle brackets represent time averaging [41]. The electrostrictive force-density of Eq.(20) has two distinguishing features: (i) the force-density is independent of the polarization state of the field; (ii) for a cylindrically symmetric intensity profile of the E-field, the force-density is cylindrically symmetric.

A key question now arises as to whether the Lorentz force-density $F_L$ of Eq.(3) or the Einstein-Laub force-density $F_{EL}$ of Eq.(11) provides a better model for electrostrictive phenomena in non-magnetic dielectrics for which we hereinafter set $\rho_{free}=0$, $J_{free}=0$, and $M=0$. Our arguments in the preceding sections reveal that the time-averaged Einstein-Laub force density is in fact nearly identical to the standard model result [to within a numerical factor of $(n^2+2)/3$], whereas the time-averaged Lorentz force does not yield a cylindrically symmetric force-density for an intensity profile of the same symmetry, and that the force-density further depends on the polarization state. This suggests that excitation of acoustic waves in a medium by a laser beam could serve as a useful experimental test of the Lorentz versus Einstein-Laub theories. The fact that, in its general features, the Einstein-Laub formulation agrees with the standard model of electrostriction is a powerful argument in its favor given the success of the standard model in describing the nonlinear optical phenomena of electorstrictive self-focusing and stimulated Brillouin scattering [40-43]. However, the possibility still exists that the standard theory is lacking and that the Lorentz theory is indeed correct. By highlighting the differences between the force densities predicted by the two theories, perhaps the numerical simulations reported in the present paper could suggest to the experimentalist the design of some possible diagnostic tests.

**7. Gaussian beam propagating through a water droplet**. Figure 7 shows a focused laser beam having $\lambda_o = 532$ nm and amplitude FWHM = 2.0 µm, illuminating a spherical water droplet of refractive index $n = 1.35$ and diameter $d = 3$ µm. The beam is linearly polarized along the y-axis, and the droplet is centered at the origin of coordinates, its upper surface being 0.2 µm below the focal plane of the lens.

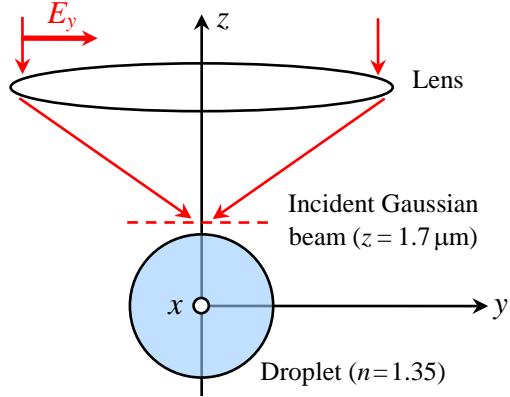

**Fig.7 (Color online)**. A green laser beam ($\lambda_o = 532$ nm), linearly polarized along the y-axis, is focused above a 3 µm-diameter water droplet. The focused spot's FWHM is 2.0 µm. The droplet is centered at the origin, with the focal plane of the lens being 0.2 µm above its upper surface.

The focused beam's cross-sectional profile in the xy-plane is Gaussian, and the source plane is located at $z = 1.7$ µm, which is 0.2 µm above the droplet's upper surface. The propagation is along the negative z-axis. In the absence of the droplet, Fig.8 shows three cross-sectional plots of $E_y$, i.e., the E-field amplitude in the xy, xz, and yz planes. These $E_y$ plots are snapshots at a particular time (long enough for the FDTD solution to have become time-harmonic). The overall propagation distance in Fig.8 being less than the Rayleigh range of the focused spot, the beam remains essentially collimated as it propagates downward.

Figure 9 shows cross-sectional profiles of the field amplitude $E_y$ in the $xy$, $xz$, and $yz$ planes in the presence of the water droplet. The droplet now acts as a second lens to further focus the incident beam into a smaller spot below the droplet. Reflection and refraction at the air-water interface as the beam enters and exits the droplet are fully accounted for in these simulations.

The force-density plots in Figs. 10, 11 and 12 represent local time averages. The $F_x$, $F_y$ and $F_z$ force-density distributions are shown in the $xy$, $xz$, and $yz$-planes through the center of the droplet. The $F_x$ and $F_y$ profiles have a symmetry plane where they vanish, so the corresponding plots are not included in Figs. 10 and 11. Within the droplet, the $F_x$ and $F_z$ force components have similar profiles in the two formulations. However, $F_y$ shows two distinct behaviors, tending to compress the droplet in the Einstein-Laub formulation (left panel in Fig. 11), and to stretch it in the Lorentz formulation (Fig. 11, right panel). It is this feature of the $F_y$ component of the force-density distribution that must be probed experimentally in order to decide between the Einstein-Laub and Lorentz formulations.

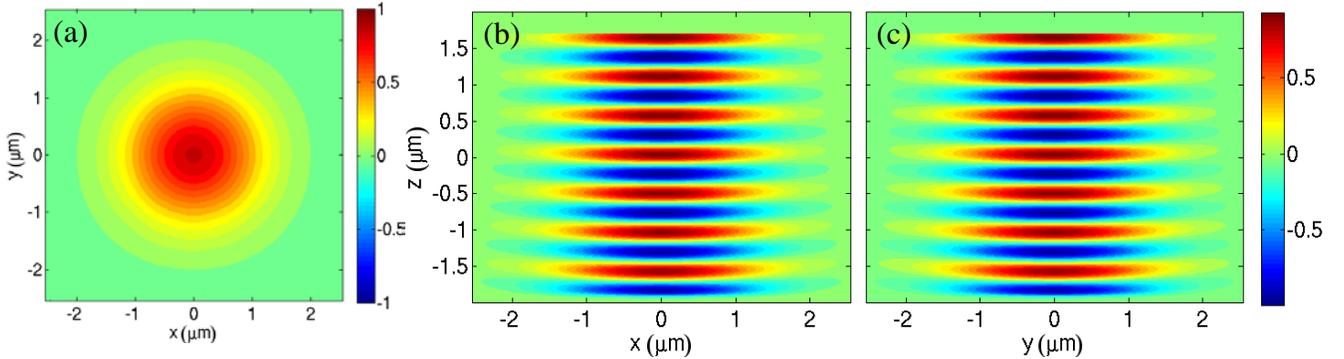

**Fig. 8 (Color online).** Cross-sectional plots of the $E_y$ field amplitude in the absence of the droplet. Frames (a) to (c) show $E_y$ in the central $xy$, $xz$, and $yz$-planes, respectively. The Gaussian beam ($\lambda_o = 532$ nm) is initiated at $z = 1.7\,\mu$m and propagates downward, along the negative $z$-axis. The integral of the $z$-component of the Poynting vector over the $xy$-plane is $P_{\text{inc}} = 3.0 \times 10^{-15}$ W. This arbitrary optical power is used for numerical convenience; all computed EM forces must be normalized by $P_{\text{inc}}$ to yield the force per watt of incident optical power.

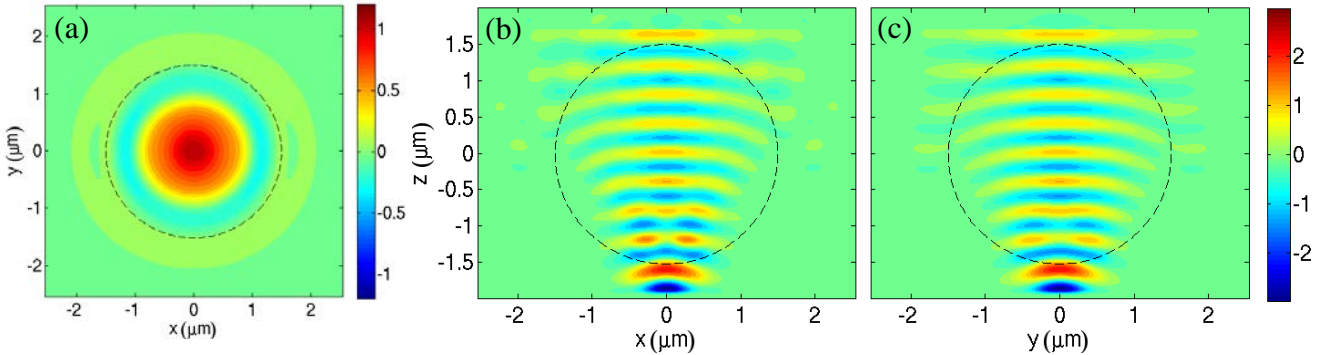

**Fig. 9 (Color online).** Cross-sectional plots of the $E_y$ field amplitude in the presence of the droplet ($n = 1.35$, $d = 3\,\mu$m). The Gaussian beam is initiated at $z = 1.7\,\mu$m and propagates downward, along the negative $z$-axis. The $xy$ cross-sectional plot in (a) represents the central plane of the droplet, where $z = 0$. Similarly, the $xz$ cross-sectional plot in (b) corresponds to the central plane where $y = 0$, and the $yz$ cross-sectional plot in (c) represents the central plane located at $x = 0$.

Figure 10 shows plots of $F_x$ in two cross-sectional planes: top row: $xy$-plane, bottom row: $xz$-plane. The symmetry of the problem dictates that $F_x$ vanish in the central $yz$-plane. The left-hand column in Fig. 10 corresponds to the Einstein-Laub formulation, while the right-hand column represents the Lorentz formulation. In the case of $F_x$, there is not much difference between the two formulations: both predict a force that tends to compress the droplet toward the central $yz$-plane. The integrated force on each hemisphere in both formulations is $\pm 6.97 \times 10^{-18}\,\mu\text{N}$. Normalization by the incident optical power $P_{\text{inc}}$ yields a force of $\pm 2.3\,\text{nN}$ per watt on each hemisphere.

Figure 11 shows plots of $F_y$ in two cross-sectional planes: top row: $xy$-plane, bottom row: $yz$-plane. The symmetry of the problem dictates that $F_y$ vanish in the central $xz$-plane. The left-hand column corresponds to the Einstein-Laub formulation, while the right-hand column represents the Lorentz formulation. For this force component, there is a substantial difference between the two formulations: While $F_{y(\text{EL})}$ tends to compress the droplet toward the central $xz$-plane, $F_{y(\text{L})}$ wants to stretch the liquid away from that central plane. The integrated force on each hemisphere for the Einstein-Laub formulation is $\pm 2.3\,\text{nN}$ per watt of incident optical power; the corresponding Lorentz force is $\pm 2.96\,\text{nN}$ per watt.

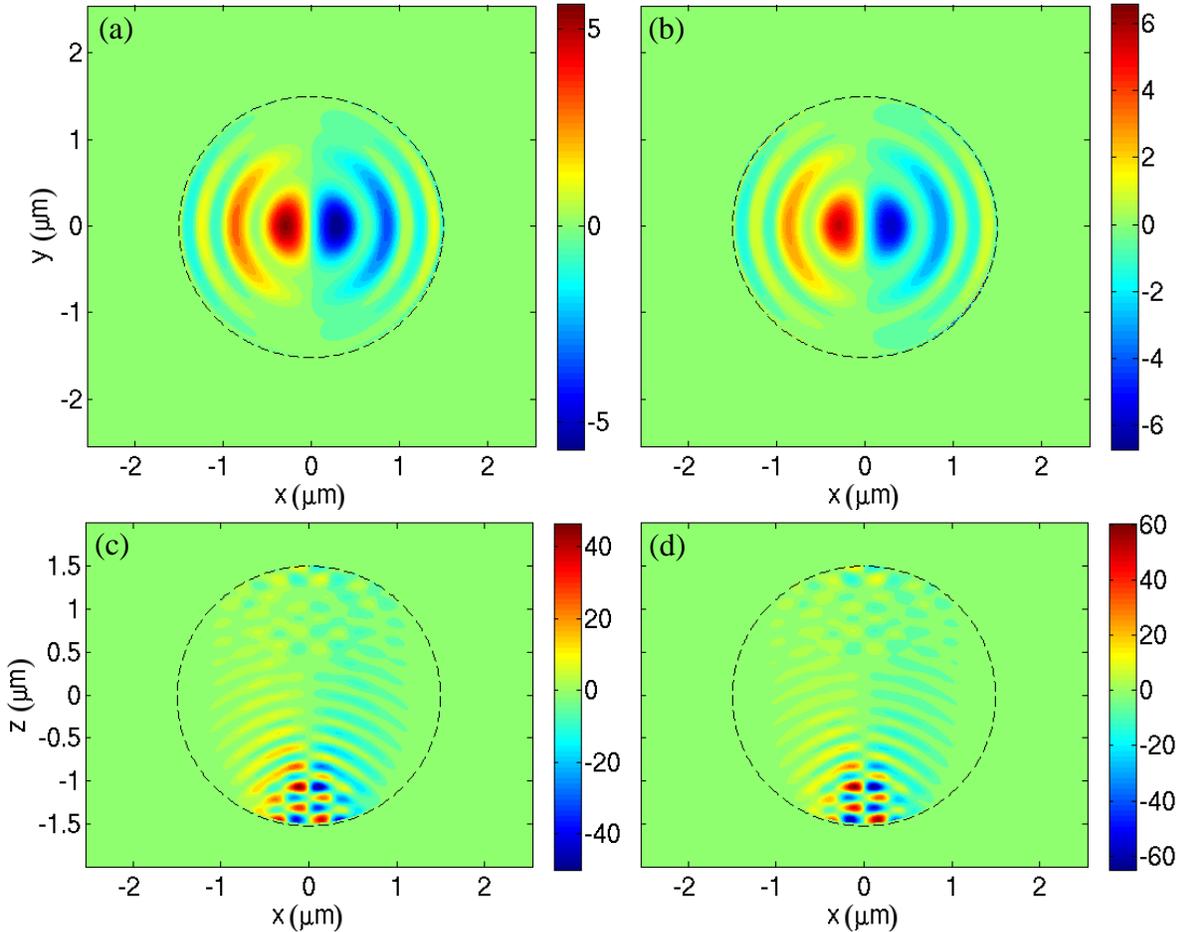

**Fig. 10 (Color online).** Plots of $F_x$ in two cross-sectional planes: top row: $xy$-plane, bottom row: $xz$-plane. The symmetry of the problem is such that $F_x$ vanishes in the central $yz$-plane. The left-hand column corresponds to the Einstein-Laub formulation, while the right-hand column represents the Lorentz formulation. In both formulations $F_x$ tends to compress the droplet toward the central $yz$-plane. The color scale-bars indicate the computed force-density in $\mu\text{N/m}^3$, corresponding to an incident optical power $P_{\text{inc}} = 3.0 \times 10^{-15}\,\text{W}$.

Figure 12 shows plots of $F_z$ in three cross-sectional planes: top row: $xy$-plane, middle row: $xz$-plane, bottom row: $yz$-plane. The left-hand column corresponds to the Einstein-Laub formulation, while the right-hand column represents the Lorentz formulation. For this force component, there is not much difference between the two formulations: both predict an $F_z$ that, on the whole, tends to push the droplet downward and away from the lens. The total integrated value of $F_z$ in both cases is $-0.53\,\text{nN}$ per watt of incident optical power. (This means that, in an inverted optical trap, the $0.14\,\text{pN}$ weight of the droplet can be balanced against a $260\,\mu\text{W}$ continuous wave (cw) focused laser beam.)

It is thus clear that the EM force distribution inside a liquid droplet is substantially different in the two formulations. If the droplet visibly deforms under a focused beam, its deformation, according to Einstein and Laub, will be axially symmetric and independent of the polarization state of the incident beam. The droplet thus assumes the shape of a prolate spheroid. In contrast, the Lorentz force tends to flatten the droplet by stretching it in the equatorial plane along the direction of incident polarization, while maintaining the vertical distance between its poles.

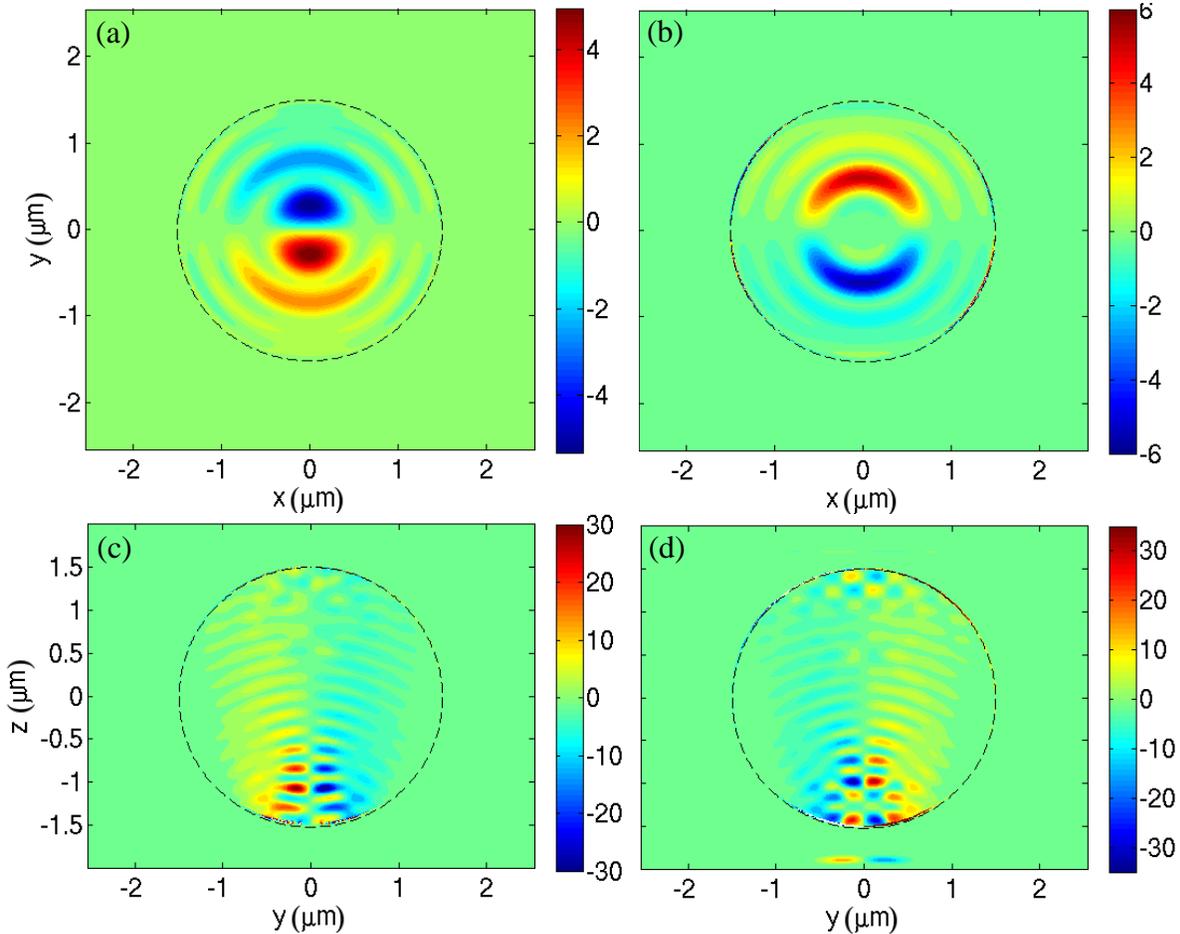

**Fig. 11 (Color online).** Plots of $F_y$ in two cross-sectional planes: top row: $xy$-plane, bottom row: $yz$-plane. The symmetry of the problem is such that $F_y$ vanishes in the central $xz$-plane. The left-hand column corresponds to the Einstein-Laub formulation, while the right-hand column represents the Lorentz formulation. While $F_{y(\text{EL})}$ tends to compress the droplet toward the central $xz$-plane, $F_{y(\text{L})}$ wants to stretch the liquid away from that central plane. The color scale-bars indicate the computed force-density in $\mu\text{N/m}^3$, corresponding to an incident optical power $P_{\text{inc}} = 3.0 \times 10^{-15}\,\text{W}$.

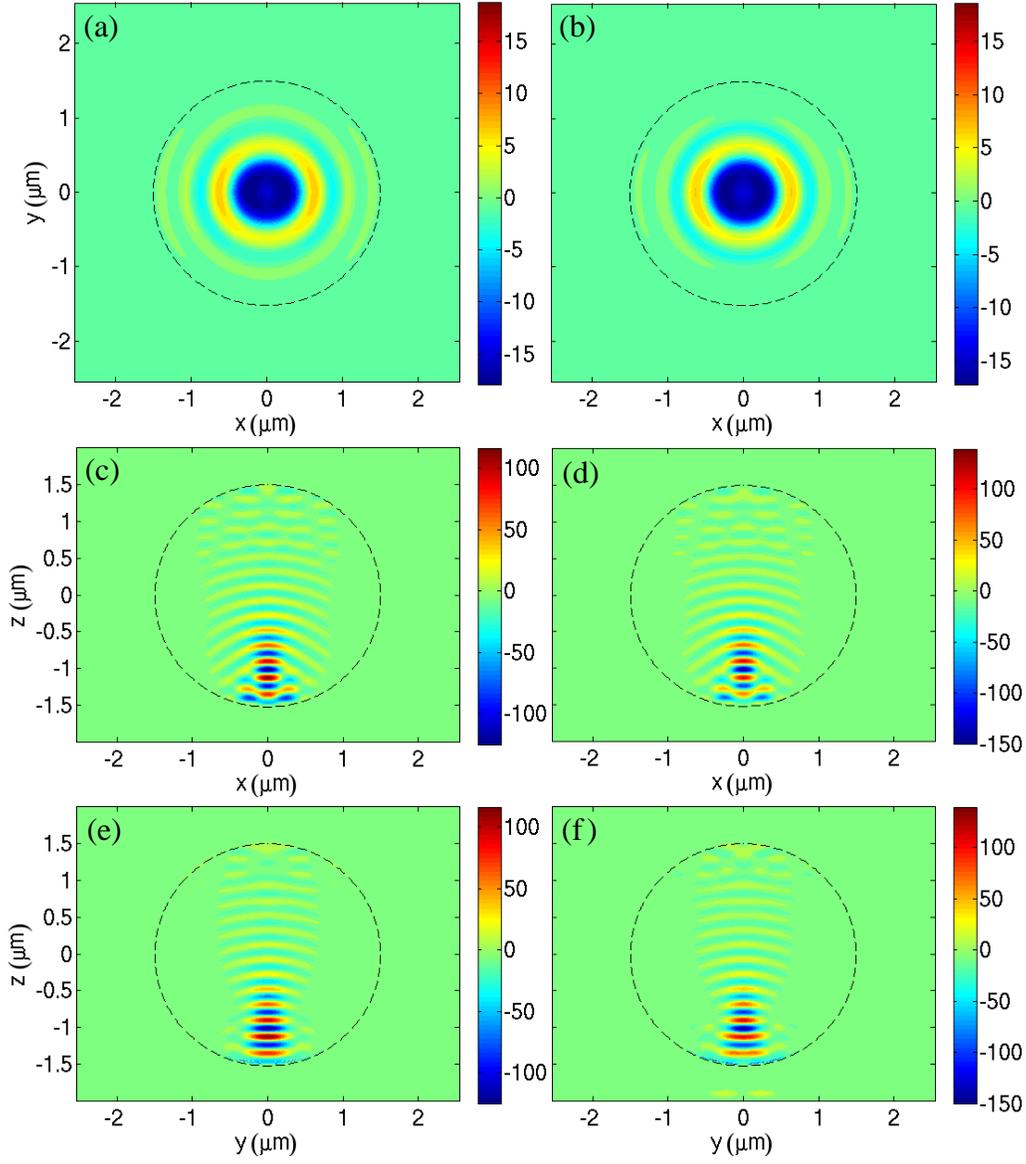

**Fig. 12 (Color online).** Plots of $F_z$ in three cross-sectional planes: top row: $xy$-plane, middle row: $xz$-plane, bottom row: $yz$-plane. The left-hand column corresponds to the Einstein-Laub formulation, while the right-hand column represents the Lorentz formulation. Both formulations predict an $F_z$ that tends to push the droplet downward. The color scale-bars indicate the computed force-density in $\mu N/m^3$, corresponding to an incident optical power $P_{\text{inc}} = 3.0 \times 10^{-15}$ W.

**8. Concluding remarks.** The force and torque exerted by EM fields on material media are intimately tied to the linear and angular momenta of the fields. Denoting the field's linear momentum density by $\boldsymbol{\rho}^{(\text{EM})}(\boldsymbol{r},t)$ and its angular momentum density by $\boldsymbol{\mathcal{L}}^{(\text{EM})}(\boldsymbol{r},t) = \boldsymbol{r} \times \boldsymbol{\rho}^{(\text{EM})}(\boldsymbol{r},t)$, conservation of momentum demands that the following identities remain valid at all times $t$:

$$\iiint_{-\infty}^{\infty} \boldsymbol{F}(\boldsymbol{r},t)\,\mathrm{d}x\mathrm{d}y\mathrm{d}z + \frac{\mathrm{d}}{\mathrm{d}t}\iiint_{-\infty}^{\infty} \boldsymbol{\rho}^{(\text{EM})}(\boldsymbol{r},t)\,\mathrm{d}x\mathrm{d}y\mathrm{d}z = 0, \tag{21a}$$

$$\iiint_{-\infty}^{\infty} \boldsymbol{T}(\boldsymbol{r},t)\,\mathrm{d}x\mathrm{d}y\mathrm{d}z + \frac{\mathrm{d}}{\mathrm{d}t}\iiint_{-\infty}^{\infty} \boldsymbol{\mathcal{L}}^{(\mathrm{EM})}(\boldsymbol{r},t)\,\mathrm{d}x\mathrm{d}y\mathrm{d}z = 0. \tag{21b}$$

Thus, in a closed system, any momentum (linear or angular) that leaves the EM field will enter the material media by way of the force and torque exerted by the fields on the media. Similarly, any momentum appearing in the EM field comes at the expense of the system's mechanical momenta (linear or angular), again through the action of EM force and torque. It can be shown that, in the case of the Lorentz law, the identities in Eq.(21) are satisfied with $\boldsymbol{F}$, $\boldsymbol{T}$, and $\boldsymbol{\wp}^{(\mathrm{EM})}$ given, respectively, by Eqs.(3), (4) and (7). Similarly, in the Einstein-Laub formulation, the identities in Eq.(21) are satisfied when the EM force, torque, and momentum are given by Eqs.(11), (12) and (15). It must be pointed out that the aforementioned EM angular momentum density expression, $\boldsymbol{\mathcal{L}}^{(\mathrm{EM})}(\boldsymbol{r},t)=\boldsymbol{r}\times\boldsymbol{\wp}^{(\mathrm{EM})}(\boldsymbol{r},t)$, is completely general, covering both the spin and orbital angular momenta of the field.

While the Lorentz formalism requires the notions of hidden energy and hidden momentum in order to comply with special relativity and with the conservation laws, the Einstein-Laub formulation remains compatible with other physical principles *without* the need for such hidden entities in conjunction with magnetic materials [7-22]. The EM momentum density in the Einstein-Laub theory, $\boldsymbol{\wp}^{(\mathrm{EM})}=\boldsymbol{E}\times\boldsymbol{H}/c^2$, is the well-known Abraham momentum density [44], which is also the momentum predicted by the Balazs thought experiment [45]. The other well-known expression for the momentum density of the EM fields is that due to Minkowski, $\boldsymbol{\wp}^{(\mathrm{EM})}=\boldsymbol{D}\times\boldsymbol{B}$, [46], which is not relevant to the present discussion; we have discussed the significance of Minkowski's momentum and its role in radiation pressure problems elsewhere [47].

Since hidden momentum is not a measurable entity, it is impossible to decide between the two theories on the basis of hidden momentum. It is also known that, once the contribution of hidden momentum is properly removed from the Lorentz law, the *total* force and *total* torque exerted by EM fields on any isolated object will turn out to be precisely the same [5,25]. Any measurable differences between the two formulations must therefore be sought in the *distributions* of EM force and torque within material media. In the examples presented in Sections 4, 5 and 7, we focused our attention exclusively on non-magnetic materials, where small deformations of the host medium (induced by radiation pressure acting on electric dipoles) offer the possibility of telling the two theories apart. Similar differences also exist between predicted force-density distributions inside magnetic materials, albeit at frequencies well below the optical regime, where ordinary magnetic materials exhibit significant susceptibilities. The goal of the present paper has been to draw attention to the existence of such differences, in the hope of encouraging the experimentalist to take a closer look at the internal distributions of EM force and torque.

**Acknowledgement**. We are grateful to Kishan Dholakia and Michael Mazilu of the University of St Andrews for helpful discussions. We also thank the anonymous referee who pointed out the relationship between our work and the works reported in [29] and [33].